\documentclass{article}
\usepackage{amsfonts}
\usepackage{amsmath}

\input{tcilatex}
\begin{document}

\title{Time-Fractional KdV Equation: Formulation and Solution using
Variational Methods}
\author{S. A. El-Wakil, E. M. Abulwafa, \and M. A. Zahran and A. A. Mahmoud 
\\
Theoretical Physics Research Group, Physics Department, \\
Faculty of Science, Mansoura University, Mansoura 35516, Egypt}
\date{}
\maketitle

\begin{abstract}
In this work, the semi-inverse method has been used to derive the Lagrangian
of the Korteweg-de Vries (KdV) equation. Then, the time operator of the
Lagrangian of the KdV equation has been transformed into fractional domain
in terms of the left-Riemann-Liouville fractional differential operator. The
variational of the functional of this Lagrangian leads neatly to
Euler-Lagrange equation. Via Agrawal's method, one can easily derive the
time-fractional KdV equation from this Euler-Lagrange equation. Remarkably,
the time-fractional term in the resulting KdV equation is obtained in Riesz
fractional derivative in a direct manner. As a second step, the derived
time-fractional KdV equation is solved using He's variational-iteration
method. The calculations are carried out using initial condition depends on
the nonlinear and dispersion coefficients of the KdV equation. We remark
that more pronounced effects and deeper insight into the formation and
properties of the resulting solitary wave by additionally considering the
fractional order derivative beside the nonlinearity and dispersion terms.

\begin{description}
\item $\boldsymbol{Keywords:}$ Riemann-Liouvulle fractional differential
operator; Euler-Lagrange equation; Riesz fractional derivative; fractional
KdV equation; He's variational-iteration method; solitary wave.

\item \textbf{PACS:} 05.45.Df, 05.30.Pr
\end{description}
\end{abstract}

\section{Introduction}

Most methods of classical mechanics deal with conservative systems, while
almost of the processes observed in the physical real world are
non-conservative. So, if the Lagrangian of the system is constructed using
fractional derivatives, the resulting equations of motion can be
non-conservative [1, 2]. Therefore in many cases, the real physical
processes could be modeled in a realable manner using fractional
differential equations rather integer-order equations [3]. Bateman [4] used
a Lagrangian, which leaded to an Euler-Lagrange equation that is, in some
sense, equivalent to the desired equation of motion. As a further step,
Riewe [1, 2] formulated a version of the Euler-Lagrange equation for
problems of calculus of variation with fractional derivatives. Recently,
further studies concerning the fractional Euler-Lagrange equations can be
found in the works of Agrawal and coworkers [5-8], Baleanu and coworkers
[9-15], Tarasov and Zaslavsky [16, 17] and others [18-22].

In last decades, much interest was devoted to apply fractional calculus to
almost every field of science, engineering and mathematics [5-22]. The
awareness of the importance of this type of equation has grown continuously
include for viscoelasticity and rheology, image processing, mechanics,
mechatronics, physics, and control theory, see for instance [23-25].

On the other hand, the Korteweg-de Vries (KdV) equation has been used to
describe a wide range of physics phenomena as a model for the evolution and
interaction of nonlinear waves. It was first derived as an evolution
equation that governing a one dimensional, small amplitude, long surface
gravity waves propagating in a shallow channel of water [26]. Subsequently
the KdV equation has arisen in a number of other physical contexts as
collision-free hydro-magnetic waves, stratified internal waves, ion-acoustic
waves, plasma physics, lattice dynamics, etc [27]. Certain theoretical
physics phenomena in the quantum mechanics domain are explained by means of
a KdV model. It is used in fluid dynamics, aerodynamics, and continuum
mechanics as a model for shock wave formation, solitons, turbulence,
boundary layer behavior, and mass transport. All of the physical phenomena
may be considered as non-conservative, so they can be described using
fractional differential equations. Therefore, in this work, our motive
mainly devoted to formulate a time-fractional KdV equation version using the
Euler-Lagrange equation via what is called variational method [5-7].

It was mention that several methods have been used to solve fractional
differential equations such as for example: Laplace transformation method
[28, 29], Fourier transformation method [28, 29], the iteration method [30],
and the operational method [31]. However, most of these methods are suitable
for special types of fractional differential equations, namely the linear
with constant coefficients. However, recently there are some papers deal
with the existence and multiplicity of solution of nonlinear fractional
differential equation using techniques of nonlinear analysis such as
fixed-point theorems, Leray--Shauder theory, Adomian decomposition method
and variational-iteration method [32-38]. In this paper, the obtained
fractional KdV equation will be solved using the variational-iteration
method (VIM), firstly used by He [39-41]. In addition, we will give the
estimates of the role of fractional derivative to the nonlinear and
dispersion terms in the fractional KdV equation.

This paper is organized as follows: Section 2 is devoted to describe the
formulation of the time-fractional KdV (TFKdV) equation using the
variational Euler-Lagrange method. In section 3, the resultant TFKdV
equation is solved approximately using VIM. Section 4 contains the results
and discussion of this work.

\section{The time-fractional KdV equation}

The regular KdV equation in (1+1) dimensions is given by [26]%
\begin{equation}
\frac{\partial }{\partial t}u(x,t)+A~u(x,t)\frac{\partial }{\partial x}%
u(x,t)+B~\frac{\partial ^{3}}{\partial x^{3}}u(x,t)=0\text{,}  \tag{1}
\end{equation}%
where $u(x,t)$ is a field variable, $x\in R$ is a space coordinate in the
propagation direction of the field and $t\in T$ ($=[0,T_{0}]$) is the time
variable and $A$ and $B$ are known coefficients.

Using a potential function $v(x,t)$, where $u(x,t)=v_{x}(x,t)$, gives the
potential equation of the regular KdV equation (1) in the form

\begin{equation}
v_{xt}(x,t)+A~v_{x}(x,t)v_{xx}(x,t)+B~v_{xxxx}(x,t)=0\text{,}  \tag{2}
\end{equation}%
The subscripts denote the partial differentiation of the function with
respect to the parameter. The Lagrangian of this regular KdV equation (1)
can be defined using the semi-inverse method [42, 43] as follows:

The functional of the potential equation (2) can be represented by%
\begin{equation}
J(v)=\dint\limits_{R}dx\dint\limits_{T}dt%
\{v(x,t)[c_{1}v_{xt}(x,t)+c_{2}Av_{x}(x,t)v_{xx}(x,t)+c_{3}Bv_{xxxx}(x,t)]\}%
\text{,}  \tag{3}
\end{equation}%
where $c_{1}$, $c_{2}$ and $c_{3}$ are constants to be determined.
Integrating by parts and taking $v_{t}|_{R}=v_{x}|_{R}=v_{x}|_{T}=0$ lead to%
\begin{equation}
J(v)=\dint\limits_{R}dx\dint\limits_{T}dt\{v(x,t)[-c_{1}v_{x}(x,t)v_{t}(x,t)-%
\frac{1}{2}c_{2}Av_{x}^{3}(x,t)+c_{3}Bv_{xx}^{2}(x,t)]\}\text{.}  \tag{4}
\end{equation}

The unknown constants $c_{i}$ ($i=$ $1$, $2$, $3$) can be determined by
taking the variation of the functional (4) to make it optimal. By applying
the variation of this functional and integrating each term by parts making
use the variation optimum condition to give the following relation%
\begin{equation}
2c_{1}v_{xt}(x,t)+3c_{2}Av_{x}(x,t)v_{xx}(x,t)+2c_{3}Bv_{xxxx}(x,t)=0\text{.}
\tag{5}
\end{equation}

As well known, above equation (5) is equivalent to equation (2), so the
unknown constants becomes%
\begin{equation}
c_{1}=1/2\text{, }c_{2}=1/3\text{ and }c_{3}=1/2\text{.}  \tag{6}
\end{equation}

In addition, the functional relation given by (4) yields directly the
Lagrangian form of the regular KdV equation%
\begin{equation}
L(v_{t},~v_{x},v_{xx})=-\frac{1}{2}v_{x}(x,t)v_{t}(x,t)-\frac{1}{6}%
Av_{x}^{3}(x,t)+\frac{1}{2}Bv_{xx}^{2}(x,t)\text{.}  \tag{7}
\end{equation}

Similar to this form, the Lagrangian of the time-fractional version of the
KdV equation can be written in the form%
\begin{eqnarray}
F(_{0}D_{t}^{\alpha }v,~v_{x},v_{xx}) &=&-\frac{1}{2}[_{0}D_{t}^{\alpha
}v(x,t)]v_{x}(x,t)-\frac{1}{6}Av_{x}^{3}(x,t)+\frac{1}{2}Bv_{xx}^{2}(x,t)%
\text{, }  \notag \\
0 &\leq &\alpha <1\text{,}  \TCItag{8}
\end{eqnarray}

In (8), the fractional derivative is represented in terms of the left
Riemann-Liouville fractional derivative definition [28-30]%
\begin{equation}
_{a}D_{t}^{\alpha }f(t)=\frac{1}{\Gamma (k-\alpha )}\frac{d^{k}}{dt^{k}}%
[\int_{a}^{t}d\tau (t-\tau )^{k-\alpha -1}f(\tau )]\text{, }k-1\leq \alpha
\leq k\text{, }t\in \lbrack a,b]\text{.}  \tag{9}
\end{equation}

Thus, the functional of the TFKdV equation will take the following form%
\begin{equation}
J(v)=\dint\limits_{R}dx\dint\limits_{T}dt~F(_{0}D_{t}^{\alpha
}v,~v_{x},v_{xx})\text{,}  \tag{10}
\end{equation}%
where the time-fractional Lagrangian $F(_{0}D_{t}^{\alpha }v,~v_{x},v_{xx})$
is defined by (8).

Following Agrawal's method [5-8], the variation of functional (10) with
respect to $v(x,t)$ leads to%
\begin{equation}
\delta J(v)=\dint\limits_{R}dx\dint\limits_{T}dt~\{\frac{\partial F}{%
\partial _{0}D_{t}^{\alpha }v}\delta _{0}D_{t}^{\alpha }v+\frac{\partial F}{%
\partial v_{x}}\delta v_{x}+\frac{\partial F}{\partial v_{xx}}\delta v_{xx}\}%
\text{.}  \tag{11}
\end{equation}

As is well known, the fractional integration by parts is given by the rule
[28-30]%
\begin{equation}
\int_{a}^{b}dtf(t)_{a}D_{t}^{\alpha
}g(t)=\int_{a}^{t}dtg(t)_{t}D_{b}^{\alpha }f(t)\text{, \ \ \ }f(t)\text{, }%
g(t)\text{ }\in \lbrack a,~b]\text{.}  \tag{12}
\end{equation}%
where $_{t}D_{b}^{\alpha }$, the right Riemann-Liouville fractional
derivative defined by [30]%
\begin{equation}
_{t}D_{b}^{\alpha }f(t)=\frac{(-1)^{k}}{\Gamma (k-\alpha )}\frac{d^{k}}{%
dt^{k}}[\int_{t}^{b}d\tau (\tau -t)^{k-\alpha -1}f(\tau )]\text{, }k-1\leq
\alpha \leq k\text{, }t\in \lbrack a,b]\text{.}  \tag{13}
\end{equation}

Integrating the right-hand side of (11) by parts using formula (12) leads to%
\begin{equation}
\delta J(v)=\dint\limits_{R}dx\dint\limits_{T}dt~[_{t}D_{T_{0}}^{\alpha }(%
\frac{\partial F}{\partial _{0}D_{t}^{\alpha }v})-\frac{\partial }{\partial x%
}(\frac{\partial F}{\partial v_{x}})+\frac{\partial ^{2}}{\partial x^{2}}(%
\frac{\partial F}{\partial v_{xx}})]\delta v\text{,}  \tag{14}
\end{equation}%
noting that $\delta v|_{T}=\delta v|_{R}=\delta v_{x}|_{R}=0$.

Optimizing this variation of the functional $J(v)$, i. e; $\delta J(v)=0$,
gives the Euler-Lagrange equation for the TFKdV equation in the form%
\begin{equation}
_{t}D_{T_{0}}^{\alpha }(\frac{\partial F}{\partial _{0}D_{t}^{\alpha }v})-%
\frac{\partial }{\partial x}(\frac{\partial F}{\partial v_{x}})+\frac{%
\partial ^{2}}{\partial x^{2}}(\frac{\partial F}{\partial v_{xx}})=0\text{.}
\tag{15}
\end{equation}

Substituting the Lagrangian of the TFKdV equation (8) into this
Euler-Lagrange formula (15) gives%
\begin{equation}
-\frac{1}{2}~_{t}D_{T_{0}}^{\alpha }v_{x}(x,t)+\frac{1}{2}~_{0}D_{t}^{\alpha
}v_{x}(x,t)+Av_{x}(x,t)v_{xx}(x,t)+Bv_{xxxx}(x,t)=0\text{.}  \tag{16}
\end{equation}

Once again, substituting for the potential function, $v_{x}(x,t)=u(x,t)$,
gives the TFKdV equation for the state function $u(x,t)$ in the form%
\begin{equation}
\frac{1}{2}[_{0}D_{t}^{\alpha }u(x,t)-_{t}D_{T_{0}}^{\alpha
}u(x,t)]+A~u(x,t)~u_{x}(x,t)+B~u_{xxx}(x,t)=0\text{,}  \tag{17}
\end{equation}%
where the fractional derivatives $_{0}D_{t}^{\alpha }$ and $%
_{t}D_{T_{0}}^{\alpha }$ are, respectively the left and right
Riemann-Liouville fractional derivatives and are defined by (9) and (13).

The TFKdV equation represented in (17) can be rewritten by the formula%
\begin{equation}
\frac{1}{2}~_{0}^{R}D_{t}^{\alpha
}u(x,t)+A~u(x,t)~u_{x}(x,t)+B~u_{xxx}(x,t)=0\text{,}  \tag{18}
\end{equation}%
where the fractional operator $_{0}^{R}D_{t}^{\alpha }$ is called Riesz
fractional derivative and can be represented by [28-30]%
\begin{eqnarray}
~_{0}^{R}D_{t}^{\alpha }f(t) &=&\frac{1}{2}[_{0}D_{t}^{\alpha
}f(t)+~(-1)^{k}{}_{t}D_{T_{0}}^{\alpha }f(t)]  \notag \\
&=&\frac{1}{2}\frac{1}{\Gamma (k-\alpha )}\frac{d^{k}}{dt^{k}}%
[\int_{a}^{t}d\tau |t-\tau |^{k-\alpha -1}f(\tau )]\text{, }  \notag \\
k-1 &<&\alpha \leq k\text{, }t\in \lbrack a,b]\text{.}  \TCItag{19}
\end{eqnarray}

In this work, for the sake of completeness, we will use the
variational-iteration method (VIM) [39-41] to solve the TFKdV equation that
obtained using Euler-Lagrange variational technique.

\section{Variational-Iteration Method}

Variational-iteration method [39-41] has been used successfully to solve
different types of integer nonlinear differential equations, e. g; [40, 41].
Also, VIM is used to solve linear and nonlinear fractional differential
equations, e. g; [36-38]. Therefore, we extend this method to solve the
TFKdV-type equation.

The basic features of the VIM outlined as follows. Considering a nonlinear
partial differential equation consists of a linear part $\hat{L}U(x,t)$,
nonlinear part $\hat{N}U(x,t)$ and a free term $f(x,t)$ represented as%
\begin{equation}
\hat{L}U(x,t)+\hat{N}U(x,t)=f(x,t)\text{,}  \tag{20}
\end{equation}%
where $\hat{L}$ is the linear operator and $\hat{N}$ is the nonlinear
operator. According to the VIM, the ($n+1$)\underline{th} approximation
solution of (20) can be read using iteration correction functional as [40,
41]%
\begin{equation}
U_{n+1}(x,t)=U_{n}(x,t)+\int_{0}^{t}d\tau \lambda (\tau )[\hat{L}%
U_{n}(x,\tau )+\hat{N}\tilde{U}_{n}(x,\tau )-f(x,\tau )]\text{, }n\geq 0%
\text{,}  \tag{21}
\end{equation}%
where $\lambda (\tau )$ is a Lagrangian multiplier and $\tilde{U}_{n}(x,\tau
)$ is considered as a restricted variation function [40, 41], i. e; $\delta 
\tilde{U}_{n}(x,\tau )=0$. Extreming the variation of the correction
functional (21) leads to the Lagrangian multiplier $\lambda (\tau )$. The
initial iteration can be used as the solution of the linear part of (20) or
the initial value $U(x,0)$. As $n$ tends to infinity, the iteration leads to
the exact solution of (20), i. e;%
\begin{equation}
U(x,t)=\underset{n\rightarrow \infty }{\lim }U_{n}(x,t)\text{.}  \tag{22}
\end{equation}

\section{Time-fractional KdV equation Solution}

In this section, we will explain in some details how one can use the VIM to
solve the TFKdV equation represented by (18).

Acting from left by the fractional operator $_{0}^{R}D_{t}^{\ \alpha -1}$ on
(18) gives us%
\begin{eqnarray}
&&\frac{\partial }{\partial t}u(x,t)-~_{0}^{R}D_{t}^{\ \alpha
-1}u(x,t)|_{t=0}\frac{t^{\alpha -2}}{\Gamma (\alpha -1)}  \notag \\
&&+\ _{0}^{R}D_{t}^{\ 1-\alpha }[A~u(x,t)\frac{\partial }{\partial x}%
u(x,t)+B~\frac{\partial ^{3}}{\partial x^{3}}u(x,t)]=0\text{, }  \notag \\
0 &\leq &\alpha <1\text{, }t\in \lbrack 0,T_{0}]\text{,}  \TCItag{23}
\end{eqnarray}%
Taking into account the following fractional derivative property [28-30]%
\begin{eqnarray}
\ _{a}^{R}D_{b}^{\ \alpha }[\ _{a}^{R}D_{b}^{\ \beta }f(t)]
&=&~_{a}^{R}D_{b}^{\ \alpha +\beta }f(t)-\overset{k}{\underset{j=1}{\sum }}\
_{a}^{R}D_{b}^{\ \beta -j}f(t)|_{t=a}~\frac{(t-a)^{-\alpha -j}}{\Gamma
(1-\alpha -j)}\text{, }  \notag \\
k-1 &\leq &\beta <k\text{.}  \TCItag{24}
\end{eqnarray}%
Then, the iterative correction functional of equation (23) becomes%
\begin{eqnarray}
u_{n+1}(x,t) &=&u_{n}(x,t)+\int_{0}^{t}d\tau \lambda (\tau )\{\frac{\partial 
}{\partial \tau }u_{n}(x,\tau )  \notag \\
&&-~_{0}^{R}I_{\tau }^{1-\alpha }u_{n}(x,\tau )|_{\tau =0}\frac{\tau
^{\alpha -2}}{\Gamma (\alpha -1)}  \notag \\
&&+\ _{0}^{R}D_{\tau }^{\ 1-\alpha }[A~\tilde{u}_{n}(x,\tau )\frac{\partial 
}{\partial x}\tilde{u}_{n}(x,\tau )+B~\frac{\partial ^{3}}{\partial x^{3}}%
\tilde{u}_{n}(x,\tau )]\}\text{,}  \notag \\
\text{ }n &\geq &0\text{,}  \TCItag{25}
\end{eqnarray}%
where the function $\tilde{u}_{n}(x,t)$ is considered as a restricted
variation function, i. e; $\delta \tilde{u}_{n}(x,t)=0$. The extreme of the
variation of (25) subject to the restricted variation function
straightforwardly gives us%
\begin{eqnarray*}
\delta u_{n+1}(x,t) &=&\delta u_{n}(x,t)+\int_{0}^{t}d\tau \lambda (\tau
)~\delta \frac{\partial }{\partial \tau }u_{n}(x,\tau ) \\
&=&\delta u_{n}(x,t)+\lambda (\tau )~\delta u_{n}(x,t)-\int_{0}^{t}d\tau 
\frac{\partial }{\partial \tau }\lambda (\tau )~\delta u_{n}(x,\tau )=0\text{%
.}
\end{eqnarray*}%
This relation leads to the stationary conditions $1+\lambda (t)=0$ and $%
\frac{\partial }{\partial \tau }\lambda (\tau )=0$, which leads to the
Lagrangian multiplier as $\lambda (\tau )=-1$.

Therefore, the correction functional (25) takes the following form%
\begin{eqnarray}
u_{n+1}(x,t) &=&u_{n}(x,t)-\int_{0}^{t}d\tau \{\frac{\partial }{\partial
\tau }u_{n}(x,\tau )  \notag \\
&&-~_{0}^{R}I_{\tau }^{1-\alpha }u_{n}(x,\tau )|_{\tau =0}\frac{\tau
^{\alpha -2}}{\Gamma (\alpha -1)}  \notag \\
&&+\ _{0}^{R}D_{\tau }^{\ 1-\alpha }[A~u_{n}(x,\tau )\frac{\partial }{%
\partial x}u_{n}(x,\tau )+B~\frac{\partial ^{3}}{\partial x^{3}}u_{n}(x,\tau
)]\}\text{,}  \notag \\
\text{ }n &\geq &0\text{.}  \TCItag{26}
\end{eqnarray}%
As $\alpha <1$, the operator $_{0}^{R}D_{\tau }^{\ \alpha -1}$ reduced to
integral one. It means that, one can express the fractional operator as the
Riesz fractional integral $_{0}^{R}I_{\tau }^{1-\alpha }$ [28-30]%
\begin{equation}
\ _{0}^{R}I_{t}^{\ \alpha }f(t)=\frac{1}{2}[_{0}I_{t}^{\ \alpha }f(t)\
+~_{t}I_{b}^{\ \alpha }f(t)]=\frac{1}{2}\frac{1}{\Gamma (\alpha )}%
\int_{a}^{b}d\tau |t-\tau |^{\alpha -1}f(\tau )\text{, }\alpha >0\text{.} 
\tag{27}
\end{equation}%
Noting that $_{0}I~_{t}^{\ \alpha }f(t)$\ and $_{t}I_{b}^{\ \alpha }f(t)$
are defined as the left and right Riemann-Liouvulle fractional integrals,
respectively

In Physics, if $t$ denotes the time-variable, the right Riemann-Liouville
fractional derivative is interpreted as a future state of the process. For
this reason, the right-derivative is usually neglected in applications, when
the present state of the process does not depend on the results of the
future development [5]. Therefore, the right-derivative is used equal to
zero in the following calculations.

The zero order correction of the solution can be taken as the initial value
of the state variable, which is taken in this case as%
\begin{equation}
u_{0}(x,t)=u(x,0)=A_{0}\sec \text{h}^{2}(cx)\text{.}  \tag{28}
\end{equation}%
where $A_{0}$ and $c$ are constants.

Substituting this zero order approximation into (27) and using the
definition of the fractional derivative (19) lead to the first order
approximation as%
\begin{eqnarray}
u_{1}(x,t) &=&A_{0}\sec \text{h}^{2}(cx)  \notag \\
&&+2A_{0}c~\sinh (cx)~\sec \text{h}^{3}(cx)[4c^{2}B  \notag \\
&&+(A_{0}A-12c^{2}B)\sec \text{h}^{2}(cx)]\frac{t~^{\alpha }}{\Gamma (\alpha
+1)}\text{.}  \TCItag{29}
\end{eqnarray}

Substituting this zero order approximation into (26) and using the
definition of the fractional derivative (19) lead to the first order
approximation as%
\begin{eqnarray}
u_{2}(x,t) &=&A_{0}\sec \text{h}^{2}(cx)  \notag \\
&&+A_{0}c~\sinh (cx)\sec \text{h}^{3}(cx)  \notag \\
&&\ast \lbrack 4c^{2}B+(A_{0}A-12c^{2}B)\sec \text{h}^{2}(cx)]\frac{%
t^{\alpha }}{\Gamma (\alpha +1)}  \notag \\
&&+A_{0}c^{2}\sec \text{h}^{2}(cx)  \notag \\
&&\ast \lbrack 16c^{4}B^{2}+8c^{2}B(5A_{0}A-63c^{2}B)\sec \text{h}^{2}(cx) 
\notag \\
&&+(3A_{0}^{2}A^{2}-176A_{0}c^{2}AB+168c^{4}B^{2})\sec \text{h}^{4}(cx) 
\notag \\
&&-\frac{7}{2}(A_{0}^{2}A^{2}-42A_{0}c^{2}AB+360c^{4}B^{2})\sec \text{h}%
^{6}(cx)]\frac{t^{2\alpha }}{\Gamma (2\alpha +1)}  \notag \\
&&+A_{0}^{2}c^{3}\sinh (cx)~\sec \text{h}^{5}(cx)  \notag \\
&&\ast \{4c^{2}B[4c^{2}B+3(A_{0}A-14c^{2}B)\sec \text{h}^{2}(cx)]  \notag \\
&&+2(A_{0}^{2}A^{2}-32A_{0}c^{2}AB+240c^{4}B^{2})\sec \text{h}^{4}(cx) 
\notag \\
&&-\frac{5}{2}(A_{0}^{2}A^{2}-24A_{0}c^{2}AB+144c^{4}B^{2})\sec \text{h}%
^{6}(cx)\}  \notag \\
&&\ast \frac{\Gamma (2\alpha +1)}{[\Gamma (\alpha +1)]^{2}}\frac{t^{3\alpha }%
}{\Gamma (3\alpha +1)}.  \TCItag{30}
\end{eqnarray}

The higher order approximations can be calculated using the Maple or the
Mathematica package to the appropriate order where the infinite
approximation leads to the exact solution.

\section{Results and Discussion}

It is well-known that the nonlinear partial differential equations known as
evolution equations possess a special type of elementary solution. These
solutions known as solitons have the form of localized waves that conserve
their properties even after interaction among them, and then act somewhat
like particles. And although the balancing scenario between nonlinearity and
dispersion is one of the most common ways for solitary waves to occur, they
can also arise as a consequence of other balancing acts. Thus our motive in
this work is to study, in some details, the effects of fractional order
derivative on the structure and propagation of the resulting solitary waves
obtained from TFKdV.

In order to achieve this target, the semi-inverse method [42, 43] is used to
derive the Lagrangian of the regular KdV equation. The Lagrangian of the
TFKdV equation is then taken in a similar form with the
left-Riemann-Liouville derivative [28-30] for the time-fractional operator.
This Lagrangain form is our roadmap to find the Euler-Lagrange equation via
the variational method [5-10] that is finally applied to derive the TFKdV
equation. Surprisingly, our resulting TFKdV equation is expressed directly
in terms of Riesz differential form [28-30]. Further, the VIM suggested by
He [39-41] is employed to solve the obtained TFKdV equation.

Actually, with the adding of the Maple, we continue our calculations till
the forth order iteration of the VIM. Therefore, our approximate
calculations are carried out concerning the solution of the TFKdV equation
taking into account the values of the coefficients of nonlinearity ($A=-6$)
and dispersion ($B=1$). Noting that, the initial value that assumed for the
solution of all cases is taken as $A_{0}~\sec $h$^{2}(cx)$ where the
amplitude $A_{0}$ to be unity and the constant ($c$) assumed as $%
B/(4A)=-1/24 $. Because our main objective of the present paper is to
explore the effect of the fractional order derivative, therefore our
solution is calculated for some interesting values namely $\alpha =1$, $3/4$%
, $1/2$, $1/3$, and $1/4$.

In addition, 3-dimensional representation of the solution of the TFKdV
equation with space and time for different values of the fractional order ($%
\alpha $) is depicted in Fig (1). Interestingly, the solution $u(x.t)$ is
still a single soliton solution for all values of the fractional parameter.
This means that, the balancing scenario between nonlinearity and dispersion
is still valid. However, the structure (amplitude and width) of the soliton
has been changed.

Figure (2) gives a good impression about the change of amplitude and width
of the soliton due to the variation of the fractional power order.
Remarkably, both 2- and 3-dimensional graphs depicted the behavior of the
solution $u(x.t)$ against the space $x$ at time $t$ corresponding to
different values of the fractional order $\alpha $. The behavior shows that
the increasing of the fractional parameter $\alpha $ increasing both the
height and the width of the solitary wave solution. This means that, the
fractional parameter can be used to modify the shape of the solitary wave
without change of the nonlinearity and the dispersion effects in the medium.

Figure (3) devoted to study the relation between the amplitude of the
soliton and the fractional order $\alpha $ at different time values. These
figures show that, at the same time, the increasing of the fractional
parameter $\alpha $ increases the amplitude of the solitary wave to some
value of then the amplitude decreases with increasing of $\alpha $, i. e;
with values very close to unit.

These results show that the fractional order of the TFKdV equation can be
used to estimate the effect of the higher order of the dispersion of the
regular KdV equation to increase the amplitude of the soliton. Also, the
results could be modified in obvious manner if the fractional order has been
applied also to the space differentiation.

$\mathbf{Figure~Captions}$

Fig. 1: The distribution function $u(x,t)$ as a 3-dimensions graph for
different values of the fractional order ($\alpha $).

Fig. 2: The distribution function $u(x,t)$ as a function of space $x$ at
time $t$ for different values of the fractional order ($\alpha $): (a)
3-dimensions graph and (b) 2-dimensions graph.

Fig. 3: The amplitude of the distribution function $u(0,t)$ as a function of
the fractional order ($\alpha $) at different time values: (a) 3-dimensions
graph and (b) 2-dimensions graph.

\end{document}